\documentclass[aps,twocolumn,floatfix,showpacs,amssymb]{revtex4}
\usepackage{graphicx,bm}
\usepackage{hyperref}
\usepackage{amsmath}
\usepackage{mathbbol}
\usepackage{color}

\begin{document}

\title{A Pure Confinement Induced Trimer in One-Dimensional Atomic Waveguides}

\author{Ludovic Pricoupenko}

\affiliation
{
Laboratoire de Physique Th\'{e}orique de la Mati\`{e}re Condens\'{e}e, Sorbonne Universit\'{e},  CNRS UMR 7600, F-75005, Paris, France.
}
\date{\today}


\begin{abstract}
Shallow trimers composed of three bosonic atoms in one-dimensional harmonic waveguides are studied in the vicinity of a Feshbach resonance. It is shown that for arbitrarily large values of the one-dimensional scattering length, an excited trimer branch exists in coexistence with the dimer and the trimer of the Lieb-Liniger model. 
 \end{abstract}

\maketitle

In current ultracold atoms experiments, low temperatures and high aspect ratio for anisotropic traps permit one to explore properties of quantum gases in one-dimensional (1D) and two-dimensional (2D) geometries \cite{QGLD}. Moreover the use of the magnetic Feshbach Resonance (FR) mechanism is an efficient and precise way to tune the three-dimensional (3D) scattering length and thus the pairwise effective interaction between atoms. Due to the large separation of scale between the range of interatomic forces (typically of few nanometers) and the characteristics lengths in trapped ultracold gases, universal expressions of the 1D (or 2D) effective interaction in harmonic waveguides can be deduced from the 3D interaction \cite{Ols98,Pet00,Pet01,Ber03}. This led to the discovery and the achievement of Confinement Induced Resonances \cite{Hal10}. These findings paved the way of substantial progress in the studies of highly correlated many-body systems in low dimensional systems \cite{Blo08,Mor03} and have sparked intensive studies on integrable systems based on the (purely 1D) Lieb Liniger model \cite{Lie63,McG64,Gua13}. A beautiful example is given by the Tonks and Super-Tonks Girardeau gas \cite{Gir60,Kin04a,Par04,Ast05,Hal09}.  Few-body systems in such configurations attract also considerable interest \cite{Kar09,Pet12,Mor15,Kri16a}. In this context, the fate of Efimov states achieved in ultracold experiments \cite{Kra06,Fer10} as a function of the strength of an external 1D or 2D harmonic confinement is an issue raised some years ago which is not fully explored yet \cite{Mor05,Lev14}. 

In this Letter, the three-boson problem in a 1D harmonic waveguide is explored by using a description of the low-energy interaction processes for ultracold atoms including the Feshbach mechanism. It is shown that, as a result of the virtual excitations in the transverse modes, an excited trimer still exists in the quasi-1D limit at the threshold of the 1D dimer,  in coexistence with the Mc Guire trimer of the Lieb-Liniger model. This result exemplifies the subtle difference between the quasi-1D and the purely 1D physics, in a limit where naively both regimes are usually considered to coincide. The two lowest trimer branches are obtained analytically near the threshold dimer and are computed numerically otherwise. They are continuously connected with the lowest Efimov states found in absence of the atomic waveguide.

\begin{figure}
\includegraphics[width=8cm]{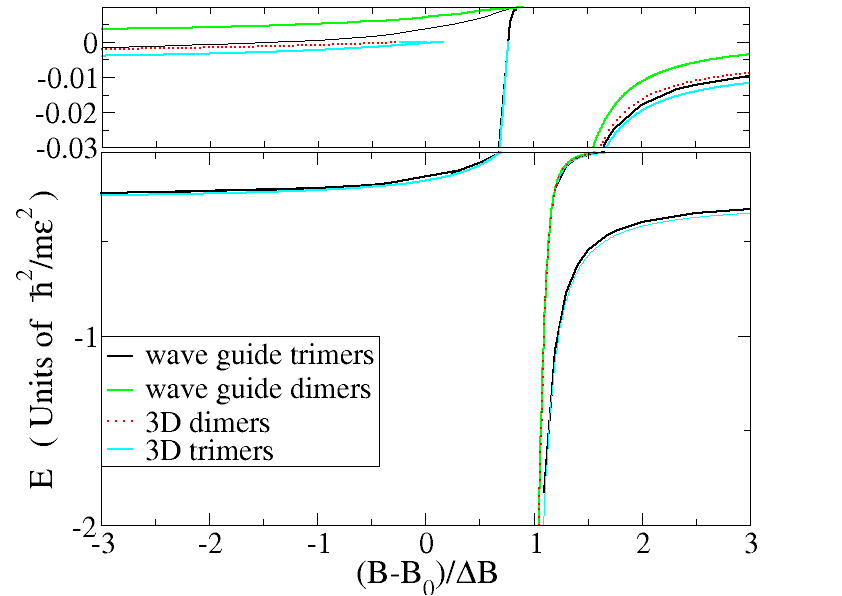}
\caption{Black solid line: trimer spectrum in the waveguide as a function of the external magnetic field obtained for a ratio ${a_\perp/\epsilon=20}$ in the case of the broad resonance at $B_0\sim-12$~G with $\Delta B\sim28.7$~G \cite{Jon10};  green solid line: dimer spectrum in the waveguide; red doted line: 3D dimer spectrum; turquoise solid line: 3D Efimov spectrum.}
\label{fig:Cesium}
\end{figure}

The 1D waveguide is modeled  by an isotropic 2D harmonic trap of frequency $\omega$. For a given atomic species of mass $m$ the characteristic length is 
\begin{equation}
a_\perp = \sqrt{\frac{2\hbar}{m \omega}} .
\label{eq:def_aperp}
\end{equation}
In this geometry, the single particle eigenstates are conveniently labeled by using the cylindrical quantum numbers. In this basis, for a radial quantum number ${n}$ and an angular momentum ${m \hbar}$ along ${z}$,  the energy of the transverse harmonic oscillator is ${(2 n + |m| +1)\hbar \omega}$. 
The three-boson problem is studied in the center of mass frame and the atoms are labeled by the index ${i\in(1,2,3)}$. The associated three sets of Jacobi coordinates are denoted by
\begin{equation}
\mathbf u_k = \mathbf r_i-\mathbf r_j
\quad ; \quad 
\underline{\mathbf u}_k = \frac{2}{\sqrt{3}}\left(\frac{\mathbf r_i+ \mathbf r_j}{2} - \mathbf r_k \right)
\label{eq:Jacobi_x}
\end{equation}
where ${(i,j,k)\in (1,2,3)^3}$ are in cyclic order. The model of interaction between atoms is a two-channel model including the coherent coupling with a molecular state in a closed channel at the heart of the Feshbach mechanism. This model was previously used for the description of Efimov states and for two-body systems in atomic waveguides \cite{Jon10,Kri15}. The short range character of the interatomic forces is modeled in this model by a generic Gaussian cut-off function. For a relative momentum ${k_0}$:
\begin{equation}
\chi_\epsilon(k_0) = \exp\left(-\frac{k_0^2 \epsilon^2}{4}\right) .
\label{eq:Gaussian}
\end{equation}
The length $\epsilon$ in Eq.~\eqref{eq:Gaussian} is of the order of the range of the interatomic forces. In the numerical computations ${\epsilon=\frac{1}{2}(m C_6/\hbar^2)^{1/4}}$ where $C_6$ is the coefficient of  the $-1/r^6$ tail of the actual interatomic potential. In the case of a 1D waveguide, the quantum numbers associated with the degrees of freedom ${\mathbf v_1}$ [${\mathbf u_1}$] are ${\alpha=(\underline{n},\underline{m},\sin \theta k_z)}$ [${\beta=(n,m,\kappa_z)}$] where ${\theta=2\pi/3}$. For the degrees of freedom ${\mathbf v_2}$ [${\mathbf u_2}$] they are denoted by ${\alpha'=(\underline{n}',\underline{m},\sin \theta k_z')}$ [${\beta'=(n',m',\kappa_z')}$]. The separable interaction and Feshbach coupling take place in the $s$-wave sector of the relative particle, thus only the relative states ${m=m'=0}$ contribute in the interaction. Moreover, the angular momentum along $z$ of the atom-molecule state is a conserved quantity (${\underline{m}' = \underline{m}}$). In order to find the simplest formulation of the problem, one uses the Skorniakov Ter Martirosian (STM) equation for the atom-molecule wave function, which can be obtained  for arbitrary separable external potential \cite{Sko57}. Details of the model and of the derivation are given in the supplemental material. In the case of a 1D waveguide, it takes the form
\begin{multline}
 2 \sum_{\underline{n}'=0}^\infty \int \frac{dk'}{2\pi} 
\langle\underline{n},\underline{m},k|\mathcal{K}(E+i0^+)|\underline{n}',\underline{m},k'\rangle f(\underline{n}',\underline{m},k')\\- D(E^{\rm rel}) f(\underline{n},\underline{m},k) = 0
\label{eq:STM-1D}
\end{multline}
where  ${f(\underline{n},\underline{m},k)=F(\alpha)}$ is proportional to the atom-molecule wavefunction, the relative energy is
\begin{equation}
E^{\rm rel} = E- (2 \underline{n} + | \underline{m}| +1) \hbar \omega - \frac{3\hbar^2 k^2}{4m} 
\end{equation}
the kernel is defined by
\begin{multline}
\langle\underline{n},\underline{m},k|\mathcal{K}(E)|\underline{n}',\underline{m},k' \rangle= \frac{m}{\pi \hbar^2}
\sum_{n\ge0} \frac{\left(1-\frac{\epsilon^2}{2a_\perp^2}\right)^{n+n'}}{\left(1+\frac{\epsilon^2}{2a_\perp^2}\right)^{2+n+n'}}\\
\times\chi_\epsilon(k/2 + k') \chi_\epsilon(k + k'/2)\\
\times
\frac{d_{{(\underline{n}+\underline{m}-n)}/{2},{(\underline{n}'+\underline{m}-n')}/{2}}^{{(\underline{n}+\underline{m}+n)}/{2}}(\theta)
d_{{(\underline{n}-\underline{m}-n)}/{2},{(\underline{n}'-\underline{m}-n')}/{2}}^{{(\underline{n}-\underline{m}+n)}/{2}}(\theta)}{{2E}/{(\hbar \omega)}-4(n+\underline{n}+1)-2 |\underline{m}|-(k^2+k'\,^2+k k')a_\perp^2} 
\label{eq:kernel_1D}
\end{multline}
and $D(E)$ is essentially the inverse of the two-body transition operator [see the supplemental material]. In the kernel \eqref{eq:kernel_1D} the summation is restricted by two conditions: first, from energy conservation one has ${n + \underline{n} = n' + \underline{n}'}$ and second, only positive energy states are accessible for a single particle in the waveguide and thus ${n'\ge 0}$. In the center of mass frame, the threshold energy for the three-body continuum is at ${2 \hbar \omega}$ and we thus introduce the binding wavenumber ${q}$ defined by ${E=2 \hbar\omega - \frac{\hbar^2}{m} q^2}$. One searches for trimers in the symmetric and zero-angular momentum sector [${f(\underline{n},0,k)=f(\underline{n},0,-k)}$].
A typical numerical solution of Eq.~\eqref{eq:STM-1D} is given in Fig.~(\ref{fig:Cesium}). The STM equation has been discretized leading to a matrix with a dimension of the order of 30 000. This example shows the expected suppression of the shallow Efimov states at the resonance ${B=B_0}$ due to the breaking of the scaling invariance consecutive to the transverse trap. Moreover the 3D and quasi-1D trimer spectrum greatly differ in region where the quasi-1D dimer has a vanishing energy. In this region, there is no shallow trimer in 3D, whereas in the quasi 1D configuration, the lowest Efimov trimer is continuously connected with what will appear in what follows as the Mc Guire trimer. More interestingly, an excited trimer is found numerically up to the dimer's threshold.
In order to have more insights in this last finding, the quasi-1D limit of the STM equation is explored analytically in what follows.

In the quasi-1D regime the colliding energy is of the order of the characteristic energy of the waveguide ${(\hbar \omega)}$ and only few transverse states are populated. The two-body properties are then given in a very good approximation by the zero range limit of the two-channel model ${(\epsilon \to 0)}$. The denominator of the two-body transition operator is then by \cite{Kri15}: 
\begin{equation}
D(E) = \frac{m}{4\pi\hbar^2} \left[ \frac{1}{\tilde{a}(E)} + \frac{\zeta_H(1/2,\frac{\hbar \omega-E}{2\hbar \omega}
)}{a_\perp} \right] .
\label{eq:DZR}
\end{equation}
where ${\tilde{a}(E)}$ is the energy dependent $s$-wave scattering length of the model (see Supplemental Material). Moreover, in the quasi-1D limit the colliding energy for two atoms is near the threshold of 1D free motion, $E= \hbar \omega + \frac{\hbar k^2}{m}$ where ${ka_\perp \ll 1}$. Then, Eq.~\eqref{eq:DZR} permits one to deduce the 1D scattering length 
\begin{equation}
D(E) = - \frac{m}{2\pi\hbar^2 a_\perp^2} \left( a_{\rm 1D} + \frac{1}{ik_0} \right) .
\label{eq:DZR}
\end{equation}
where \cite{Kri15}
\begin{equation}
a_{\rm 1D} = -\frac{a_\perp}{2}\left[\frac{a_\perp}{\tilde{a}(\hbar \omega)}+\zeta(1/2) \right]
\label{eq:a1D}
\end{equation}
For standard broad resonances, $\tilde{a}$ coincides with the 3D scattering length $a_{\rm 3D}$ in the low energy regime considered. One thus recovers from Eq.~\eqref{eq:a1D} the usual expression of the 1D scattering length, which is relevant for standard resonances \cite{Ols98}. For an arbitrarily small and negative 3D scattering length, the 1D scattering length is large and positive (${a_{\rm 1D}\gg a_\perp}$) the model gives the dimer of the Lieb-Liniger model with the binding wavenumber ${q_{\rm d}=1/a_{\rm 1D}}$.  
In this limit, the projection of the equation in the ground mode ${n=0}$ of the waveguide decouples from the others modes.  For convenience, one introduces  the dimensionless momentum ${u=k/q}$, the wavefunction ${\langle u | \psi \rangle = f(0,0,k)}$, the small parameter $\eta$ and the dimensionless energy $\chi$ defined by:
\begin{equation}
q^2=q_{\rm d}^2(1+\chi) \quad ; \quad \eta=(q a_\perp)^2 .
\end{equation}
The component of the STM equation in the lowest mode takes the form 
\begin{multline}
\frac{\langle u | \psi \rangle }{\sqrt{\frac{3}{4} u^2 + 1}} +
2 \int \frac{du'}{\pi} \sum_{n=0}^\infty
\frac{1}{4^n}\frac{\langle u' | \psi \rangle}{\frac{4n}{\eta} +1 +u^2+u'\,^2+uu'}\\
=\sqrt{1+\chi} \langle u | \psi \rangle .
\label{eq:STM_1D_modified}
\end{multline}
One can verify in the exact 1D limit where ${\eta=0^+ }$, that  the wavefunction of the Mc Guire trimer
\begin{equation}
\langle u | \psi_0^{(0)} \rangle= \frac{1}{u^2 + 1} 
\end{equation}
is a solution of Eq.~\eqref{eq:STM_1D_modified} with the dimensionless energy ${\chi=2}$ corresponding to the binding wavenumber ${q=2 q_{\rm d}}$. A perturbation calculation can be done in the vicinity of the dimer threshold, to obtain the first correction to the energy of the Mc Guire trimer: ${\chi =3 + \delta \chi }$ (${\delta \chi \ll 1}$). For this purpose, one introduces the operator associated with the strict 1D-STM equation for the Mc Guire trimer
\begin{equation}
\langle u | \mathcal L_0 |\psi\rangle = \frac{\langle u | \psi \rangle }{\sqrt{\frac{3}{4} u^2 + 1}} +
2 \int \frac{du'}{\pi} \frac{\langle u' | \psi \rangle}{1 +u^2+u'\,^2+uu'}
\end{equation}
In the limit ${\eta \ll 1}$, the first order perturbation in ${\eta}$ is
\begin{equation}
\langle u | \delta \mathcal L |\psi\rangle = \eta \ln \left(\frac{4}{3}\right) \int \frac{du'}{2\pi}\langle u' | \psi \rangle
\end{equation}
The first correction to the Mc Guire unperturbed wavefunction is denoted ${\langle u |\delta \psi_0\rangle}$. Expanding the STM equation \eqref{eq:STM_1D_modified} at the first order of the perturbation gives
\begin{equation}
\langle u | \mathcal L_0 | \delta \psi_0^{(0)} \rangle + \langle u | \delta \mathcal L |\delta \psi_0 \rangle = 2 \langle u | \delta \psi_0 \rangle
+ \frac{\delta \chi}{4} \langle u |  \psi_0^{(0)} \rangle
\label{eq:perturbation_psi0}
\end{equation}
Imposing the orthogonality condition  ${\langle \psi_0^{(0)} | \delta \psi_0 \rangle=0}$, Eq.~\eqref{eq:perturbation_psi0} gives
\begin{equation}
\delta \chi = 4 \frac{\langle \psi_0^{(0)} | \delta  \mathcal L |\psi_0^{(0)} \rangle}{\langle \psi_0^{(0)} |\psi_0^{(0)} \rangle} = 4 \eta \ln \left( \frac{4}{3} \right) .
\end{equation}
For this state at the lowest order ${\eta=4(a_\perp q_{\rm d})^2}$ and thus
\begin{equation}
E_t^{(0)} \sim 2 \hbar \omega - \frac{4 \hbar^2}{m a_{\rm 1D}^2} \left[ 1 +  \frac{4 a_\perp^2}{a_{\rm 1D}^2} \ln\left(\frac{4}{3}\right) \right] .
\label{eq:Efonda}
\end{equation}
\begin{figure}
\includegraphics[width=8cm]{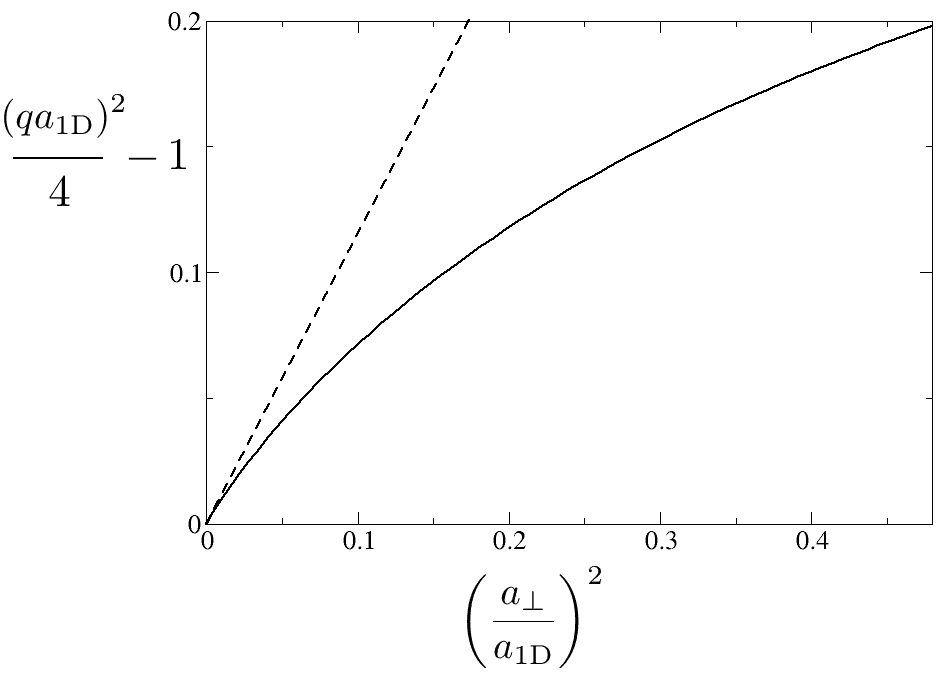}[h]
\caption{Relative difference between the lowest Quasi 1D trimer obtained from Eq.~\eqref{eq:STM_1D_modified} and the Mc Guire trimer as a function of the small parameter ${(a_\perp/a_{\rm 1D})^2}$ in the vicinity of the dimer threshold.}
\label{fig:fonda}
\end{figure}
Comparison of the result in Eq.~{eq:Efonda} with the numerical solution of the lowest state of Eq.~\eqref{eq:STM_1D_modified} is given in Fig.~(\ref{fig:fonda}). One finds also in this limit, as expected an excited trimer with a vanishing binding energy. This excited trimer was found in Ref.~\cite{Mor05}, but suspected to exist only for finite and positive values of the 1D scattering length. Instead in what follows, it is shown that this excited trimer appears at the dimer threshold ${q_{\rm d} a_\perp=0^+}$. For this purpose one can remark from Eq.~\eqref{eq:STM_1D_modified} that in the limit of a small binding energy where ${\chi \ll 1}$, the small momentum behavior of the wavefunction is proportional to $\frac{1}{\frac{3}{4}u^2+\chi}$. Then, inspired by the form of the atom-dimer wavefunction at zero energy which is a solution of Eq.~\eqref{eq:STM_1D_modified}
for ${\eta=0}$:
\begin{equation}
(2\pi) \delta(u) - \frac{4}{u^2+1},
\label{eq:atom-dimer}
\end{equation}
one expects that the wavefunction of the excited bound state is approximately given in the limit ${\eta \ll 1}$ by
\begin{equation}
\langle u | \psi_1^{(0)} \rangle =  \frac{\sqrt{3 \chi}}{\frac{3}{4}u^2+\chi} - \frac{4}{u^2+1}  .
\label{eq:psi1}
\end{equation}
In the limit where the dimensionless energy ${\chi}$ tends to zero, the first term in the parenthesis of Eq.~\eqref{eq:psi1} converges toward the delta distribution of Eq.~\eqref{eq:atom-dimer}. This ansatz coincides remarkably with the numerical solution obtained for the first excited state of Eq.~\eqref{eq:STM_1D_modified}. However, this wavefunction does not give any quantification condition for the reduced enery ${\chi}$. For this purpose, it is necessary to take into account the first correction  ${\langle k |\delta \psi_1 \rangle}$ to the function ${\langle k | \psi_1^{(0)} \rangle}$. The equation verified by ${\langle u | \delta \psi_1 \rangle}$ is obtained at the lowest order by keeping only terms linear in ${\eta}$ and ${\sqrt{\chi}}$ after injecting the ansatz of Eq.~\eqref{eq:psi1} in Eq.~\eqref{eq:STM_1D_modified} :
\begin{multline}
 \langle u | \mathcal L_0 -1 | \delta \psi_1 \rangle = \eta \ln \left(\frac{4}{3}\right) +\frac{4\sqrt{\chi}}{\sqrt{3} u^2}  \Biggl( 1
 \\
 - \frac{1}{(u^2+1)^2\sqrt{\frac{3}{4} u^2+1}}  \Biggr) .
\label{eq:deltapsi1}
\end{multline}
Introducing a small-momentum cut-off in the integral term of the left hand side of Eq.~\eqref{eq:deltapsi1}, permits one to find a solution for an arbitrary ratio ${\eta/\sqrt{\chi}}$. Nevertheless the non integral term in the left hand side of Eq.~\eqref{eq:deltapsi1} imposes a ${O(1/u^2)}$ behavior for a vanishing momentum ${u \to 0}$. This last behavior leads to a divergence of the integral term in the limit of a vanishing cut-off. One thus searches only for a specific value of the ratio  ${\eta/\sqrt{\chi}}$ such that the even solution of Eq.~\eqref{eq:deltapsi1} is regular for a vanishing momentum. Interestingly, imposing that the wavefunction of the excited state stays finite in the limit where the reduced momentum vanishes, one finds from Eq.~\eqref{eq:deltapsi1}:
\begin{equation}
4 \langle \psi_0^{(0)} | \delta \psi_1 \rangle = \eta \ln \left(\frac{4}{3}\right) +\frac{19\sqrt{\chi}}{2\sqrt{3}} .
\end{equation}
One now uses the orthogonality of the wavefunctions associated with the two bound states at the first order of perturbation ${\langle \psi_0 | \psi_1 \rangle=0}$ and thus
\begin{equation}
\langle \psi_0^{(0)}  | \psi_1^{(0)} \rangle  + \langle \delta \psi_0  | \psi_1^{(0)} \rangle  + \langle \psi_0^{(0)}  | \delta \psi_1 \rangle  =0 .
\label{eq:orthogonality_condition}
\end{equation} 
\begin{figure}
\includegraphics[width=8cm]{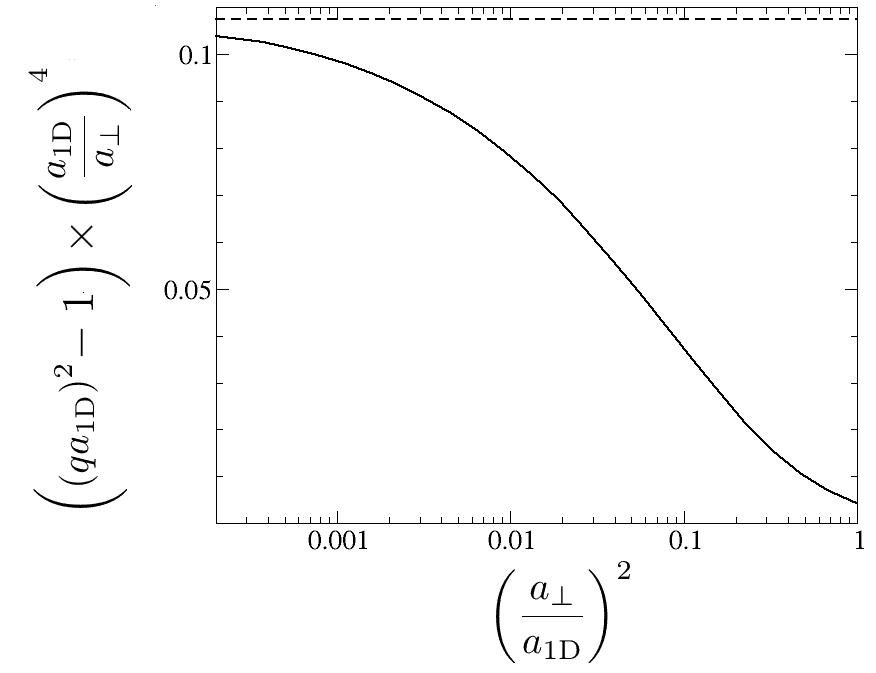}[h]
\caption{Rescaled relative binding energy of the pure confinement induced trimer as a function of the small parameter ${(a_\perp/a_{\rm 1D})^2}$ in the vicinity of the dimer threshold. Solid line: spectrum of the excited state of Eqs.~\eqref{eq:STM_1D_modified} Dashed line: analytical result of Eq.~\eqref{eq:E_t_exc} }
\label{fig:exc_trim}
\end{figure}
Multiplying Eq.\eqref{eq:perturbation_psi0} by ${\langle u |\psi_1^{(0)} \rangle}$, one obtains after integration over ${u}$:
\begin{equation}
\langle \psi_1^{(0)} | \delta  \psi_0 \rangle  = \langle \psi_1^{(0)} | \delta \mathcal L | \psi_0^{(0)} \rangle  .
\end{equation}
All the scalar products in Eq.~\eqref{eq:orthogonality_condition} can be then calculated analytically  and one obtains
\begin{equation}
\sqrt{\chi} = \frac{2 \eta}{\sqrt{3}} \ln \left( \frac{4}{3} \right) .
\end{equation}
Consequently near the threshold, the energy of the first excited trimer is given by
\begin{equation}
E_t^{(1)} = 2 \hbar \omega - \frac{\hbar^2 }{m a_{\rm 1D}^2} \left[ 1 + \frac{4 a_\perp^4 }{3 a_{\rm 1D}^4} \ln^2 \left(\frac{4}{3}\right) \right] .
\label{eq:E_t_exc}
\end{equation}
Comparison between Eq.~\eqref{eq:E_t_exc} and the numerical solution of the first (and only) excited state of Eq.~\eqref{eq:STM_1D_modified} is given in Fig.~(\ref{fig:exc_trim}).  

Equation \eqref{eq:E_t_exc} is the main result of this Letter. It proves that the threshold of the excited trimer in the quasi-1D limit coincides exactly with the dimer threshold of the Lieb Liniger model. The existence of this state is solely due to the virtual excitations in the transverse states of the 1D waveguide. As it does not exist in strickly 1D systems, it is a pure Confinement Induced Trimer. This result, not present in the Lieb Liniger model, illustrates also a breaking of integrability for quasi-1D ultracold systems.
 
Note added: while finishing the redaction of these results presented at the workshop of Ref.~\cite{Pri18}, a preprint deriving almost the same findings by using another method was sent on arxiv \cite{Nis18}.

\begin{widetext}
{\bf \large Supplemental Material: A Pure Confinement Induced Trimer in One-Dimensional Atomic Waveguides}
\end{widetext}

\section{STM equation in a separable trap}

\label{appendix:STM_separable}

The Skorniakov Ter Martirosian (STM) equation is derived below for three identical atomic bosons in a separable external potential for the model of Ref.~\cite{Jon10}.  The set of quantum numbers ${\alpha}$ associated with the degrees of freedom ${\mathbf u_k}$ (${\underline{\mathbf u}_k}$) is denoted by ${k:\alpha}$ (${\underline{k}:\alpha}$). The free Hamiltonian associated with the degree of freedom ${(\mathbf u_k)}$ ${[(\underline{\mathbf u}_k)]}$ is denoted ${\hat{h}_0^{k}}$  ${[\underline{\hat{h}}_0^{k}]}$. With the definition of the Jacobi coordinates used in this Letter, the eigenenergies for ${(k:\alpha)}$ and ${(\underline{k}:\alpha)}$ have the same expression denoted by ${\epsilon_\alpha}$:
\begin{align}
\hat{h}_0^{k} |k:\alpha \rangle = \epsilon_\alpha |k:\alpha\rangle \quad ; \quad
\underline{\hat{h}}_0^{k} |\underline{k}:\alpha \rangle = \epsilon_\alpha |\underline{k}:\alpha \rangle .
\end{align}
The formulation of the three-body problem in this two-channel model is simplified by introducing the following operator acting on the relative pair $ij$
\begin{equation}
\hat{A}_{\epsilon}^{k} = \hat{\mathbb 1}_{k} \otimes 
\langle k:\delta_\epsilon| 
\label{eq:hat_B}
\end{equation}
where ${\hat{\mathbb 1}_{k}}$ is the identity operator in the Hilbert space associated with the degrees of freedom of the atom ${k}$ and of the center of mass of the pair ${k}$ (or of one molecule). The bra ${\langle k : \delta_\epsilon|}$ denotes the bra ${\langle \delta_\epsilon|}$ that acts on the states of the relative particle for the pair $k$. This state plays the rol of a cut-off function  characterized by the short range parameter $\epsilon$ of the order of the range of the interatomic forces. In this  model, the state ${| \delta_\epsilon\rangle}$ is chosen for convenience as a Gaussian. For a pair with a relative momentum ${\mathbf k_0}$ one has:
\begin{equation}
\langle \mathbf k_0 | \delta_\epsilon \rangle\equiv \chi_\epsilon(k_0) = \exp( - k_0^2\epsilon^2/4)
\label{eq:cut-off}
\end{equation}
Using these notations, the direct interaction between two atoms is given by a separable pairwise potential characterized by the strength ${g}$:
\begin{equation}
\hat{V}^{\rm a}_{k} = g \left(\hat{A}_{\epsilon}^{k}\right)^\dagger  \hat{A}_{\epsilon}^{k} .
\label{eq:Vdirect}
\end{equation}
The coherent coupling between a pair of atoms in the open channel and a molecule is 
\begin{equation}
\hat{V}^{\rm ma}_{k} = 
\Lambda \left(\hat{A}_{\epsilon}^{k}\right) \quad ; \quad
\hat{V}^{\rm am}_{k} = \Lambda^* \left(\hat{A}_{\epsilon}^{k}\right)^\dagger .
\label{eq:Vcoupling}
\end{equation}
For an atomic pair $k$, the model leads to a two-body transition operator with a separable form \cite{Kri16b}:
\begin{equation}
\hat{T}^{\rm rel}_k = \frac{| k:\delta_\epsilon\rangle \langle k: \delta_\epsilon |}{D(E)} 
\label{eq:T-operator}
\end{equation}
where 
${
D(E) = 1/g^{\rm eff}(E) - \langle k: \delta_\epsilon |(E-\hat{h}^k_0)^{-1}|k:\delta_\epsilon \rangle
}$
and 
${
{g}^{\rm eff}(E) = g + |\Lambda|^2/(E-E_{\rm mol}) 
}$. The effective interaction strength ${{g}^{\rm eff}(E)}$ takes into account the inter-channel coupling associated with the Feshbach mechanism and can be rewritten as
\begin{equation}
{g}^{\rm eff}(E) = \frac{4\pi \hbar^2}{m} \frac{\tilde{a}(E)}{1- \sqrt{\frac{2}{\pi}} \frac{\tilde{a}(E)}{\epsilon}} 
\end{equation}
where the energy dependent scattering length
\begin{equation}
\tilde{a}(E) =  \left[\frac{1}{a}+ \frac{R^\star E \left(1-\frac{a_{\rm bg}}{a}\right)^2 }{ \frac{\hbar^2}{m}+ R^\star a_{\rm bg} E \left(1-\frac{a_{\rm bg}}{a}\right) }\right]^{-1}
\label{eq:atilde}
\end{equation}
incorporates all the relevant parameters for the description of low energy properties in the vicinity of a magnetic Feshbach Resonance. For a three-body stationary state of energy ${E}$, characterized by the three-atom state, ${|\Psi^{\rm a}\rangle}$ in the open channel associated with the state  ${|\underline{k}:\Psi^{\rm m}\rangle}$ corresponding to  one-molecule in the closed channel plus one-atom ${k}$ in the open channel, the stationary Schr\"{o}dinger equation is given by 
\begin{align}
&(E-\hat{h}_0^{k}-\underline{\hat{h}}_0^{k}) |\Psi^{\rm a}\rangle = \sum_{n} \left[ \hat{V}^{\rm a}_{n}|\Psi^{\rm a}\rangle + \hat{V}^{\rm am}_{n} |\underline{n}:\Psi^{\rm m} \rangle\right]
\label{eq:schrodi-at}\\
&(E-\underline{\hat{h}}_0^{k}-E_{\rm mol}) |\underline{k}:\Psi^{\rm m}\rangle = \hat{V}^{\rm ma}_{k} |\Psi^{\rm a}\rangle
\label{eq:schrodi-mol}
\end{align}
Combining Eqs.~\eqref{eq:schrodi-at} and \eqref{eq:schrodi-mol}, it is possible to obtain a STM equation by introducing the function ${F(\alpha)={g}^{\rm eff}(E-\epsilon_\alpha) \langle \underline{k}:\alpha | \hat{A}_{\epsilon}^{k} |\Psi^{a} \rangle}$ (which does not depend on ${k}$ due to the Bose symmetry). For three-body bound states, there is no  three-body continuum in the open channel and the STM equation takes the form
\begin{equation}
 2 \sum_{\alpha'} \langle \alpha |\mathcal{K}(E)|\alpha'\rangle F(\alpha') - D(E-\epsilon_\alpha) F(\alpha) = 0 
\label{eq:STM}
\end{equation}
The kernel in Eq.~\eqref{eq:STM} is given by
\begin{equation}
\langle \alpha |\mathcal{K}(E)|\alpha'\rangle=\sum_{\beta,\beta'} 
\frac{\langle \delta_\epsilon | \beta \rangle \langle \underline{1}:\alpha,1:\beta|\underline{2}:\alpha',2:\beta'\rangle \langle \beta' | \delta_\epsilon \rangle}{E-\epsilon_{\alpha'}-\epsilon_{\beta'}} 
\label{eq:Kernel}
\end{equation}
where the summation over the quantum numbers ${(\beta,\beta')}$ is constrained by the conservation law:
\begin{equation}
\epsilon_{\alpha'} + \epsilon_{\beta'} = \epsilon_{\alpha} + \epsilon_{\beta}
\label{eq:conserv_E}
\end{equation}
The choice of a generic Gaussian cut-off in Eq.~\eqref{eq:cut-off} gives a correct modeling of the interactions in the limit of a small harmonic confinement with respect to the potential radius (i.e. ${a_\perp \gg \epsilon}$ in the case of a 1D harmonic waveguide). This regime achieved in standard experiments, guarantees that two-body scattering in the waveguide can be indeed described by an effective model of the actual 3D interatomic forces.  

\section{Parameters of the separable two channel model}
\label{appendix:parameters}

The separable model used here gives the law verified by the 3D scattering length $a$, near a magnetic FR of width $\Delta B$, located at ${B_0}$:
\begin{equation}
a=a_{\rm bg} \left( 1 - \frac{\Delta B}{B-B_0}\right)
\label{eq:scatt-3D}
\end{equation}
For a large magnetic detuning, the inter-channel coupling can be neglected and the off-resonance (or background) 3D scattering length ${a_{\rm bg}}$ is parameterized by the strength ${g}$ of the separable potential and the cut-off $\epsilon$:
\begin{equation}
g= \frac{4\pi \hbar^2}{m} \frac{a_{\rm bg}}{1-\sqrt{\frac{2}{\pi}} \frac{a_{\rm bg}}{\epsilon}} \label{eq:g} .
\end{equation}
For simplicity, the same short-range cut-off is used in the Feshbach coupling \eqref{eq:Vcoupling}. The slope of the molecular energy $E_{\rm mol}$ as a function of the external magnetic field $B$ in the vicinity of the resonance is denoted ${\delta \mathcal M}$:
\begin{equation} 
\delta \mathcal M = \frac{\partial E_{\rm mol}}{\partial B}\biggr|_{B=B_0} .
\end{equation}
The law for the 3D scattering length in Eq.~\eqref{eq:scatt-3D} is obtained by adjusting the parameters of the model as 
\begin{align}
&{\delta \mathcal M} \Delta {B} =  \frac{4 \pi \hbar^2 \Lambda^2 a_{\rm bg}}{m g^2 } \label{eq:DeltaB} \\
&E_{\rm mol} =\delta \mathcal M (B-B_0- \Delta  B ) + \frac{\Lambda^2}{g}
\label{eq:Emol} .
\end{align}
It is also useful to introduce the the characteristic length of the Feshbach resonance ${R^\star}$ used for instance In Eq. ~\eqref{eq:atilde} \cite{Pet04b} 
\begin{equation}
R^\star = \frac{\hbar^2}{2\mu a_{\rm bg} \delta \mathcal M \Delta B}.
\label{eq:Rstar}
\end{equation}

\section{Rotation of the eigenstates in the 1D atomic waveguide}

\label{sec:hyper-rotation}
The aim of this appendix is to give a simple derivation of the scalar product between two eigenstates of the 1D waveguide, each of one being associated with a given set of Jacobi coordinates. They are needed for the evaluation of the general expression of the Kernel of the STM equation in Eq.~\eqref{eq:Kernel}. The waveguide is composed of an isotropic harmonic trap in the $(xy)$ directions and a free direction for the atomic motion along the $z$ axis. In what follows, the set of cylindrical quantum numbers ${| \underline{1}:(\underline{n},\underline{m},\underline{\kappa}_z), 1 :(n,m,\kappa_z) \rangle}$ is associated with the Jacobi coordinates ${( {\underline{\mathbf u_1},\mathbf u_1})}$ and the set ${|  \underline{2}:(\underline{n}',\underline{m}',\underline{\kappa}_z'),2 :(n',m',\kappa_z') \rangle}$ is associated with the coordinates ${( {\underline{\mathbf u_2}},\mathbf u_2)}$.

The momentum associated with the Jacobi coordinates in the direct space are defined by
\begin{equation}
\underline{\boldsymbol \pi}_k = \frac{1}{\sqrt{3}}\left( \frac{\mathbf p_i+ \mathbf p_j}{2} - \mathbf p_k \right) 
\quad ; \quad 
\boldsymbol \pi_k = \frac{\mathbf p_i-\mathbf p_j}{2} .
\label{eq:Jacobi_p}
\end{equation}
The sets of Jacobi coordinates are thus related to each other by the rotation 
\begin{equation}
\left(
\begin{array}{c}
\underline{\mathbf u}_2\\
\mathbf u_2
\end{array}
\right)
= \mathcal R(\theta)
\left(
\begin{array}{c}
\underline{\mathbf u}_1\\
\mathbf u_1
\end{array}
\right)
 \ ; \ \left(\begin{array}{c}
\underline{\boldsymbol \pi}_2\\ 
\boldsymbol \pi_2
\end{array}
\right)
= \mathcal R(\theta)
\left(
\begin{array}{c}
\underline{\boldsymbol \pi}_1\\ 
\boldsymbol \pi_1
\end{array}
\right)
\label{eq:rot_Jacobi}
\end{equation}
where ${\theta=\frac{2\pi}{3}}$ and
\begin{equation}
\mathcal R(\theta) = 
\left(
\begin{matrix}
\cos \theta & -\sin \theta \\
\sin \theta & \cos \theta
\end{matrix}
\right) . 
\end{equation}
Using the change of variables ${\underline{\kappa}_z= k_z \sin \theta}$ and ${\underline{\kappa}_z'= k_z' \sin \theta}$, one obtains the scalar product of the eigenstates of the momentum operator along $z$ from Eq.~\eqref{eq:rot_Jacobi} 
\begin{multline}
\langle \underline{1}: k_z \sin \theta , 1: \kappa_z | \underline{2}:k_z' \sin \theta, 2:  \kappa_z' \rangle\\
= \frac{(2\pi)^2}{\sin \theta} \delta(k_z+ k_z'/2 - \kappa_z')  \delta(k_z'+ k_z/2+\kappa_z)
\end{multline}
For the eigenstates of the 2D harmonic oscillator in the $(xy)$ directions, one considers the annihilation operators ${a_i^x}$ and ${a_i^y}$ (${\underline{a}_i^x}$ and ${\underline{a}_i^y}$) for the harmonic oscillator in the directions ${x}$ and  ${y}$,
associated with the Jacobi variables ${\mathbf u_i}$ (${\underline{\mathbf u}_i}$). The Jordan Wigner mapping permits one to obtain the representation of the rotation of the Jacobi coordinates for the eigenstates of the harmonic oscillator in a given direction, $x$ for example:
\begin{equation}
|\underline{2}:\underline{n}^x\,',2:n^x\,'\rangle =  R_{12}^{x}(\theta) |\underline{1}: \underline{n}^x,1:n^x \rangle
\end{equation}
with
\begin{equation}
R_{12}^{x}(\theta)= e^{\theta ((\underline{a}_1^x)^\dagger a_1^x - (a_1^x)^\dagger \underline{a}_1^x)/2}  .
\end{equation}
One recognizes a rotation operator and obtains  \cite{Lev14}
\begin{equation}
\langle \underline{1}: \underline{n}^x, 1: n^x 
| \underline{2}: \underline{n}^x\,' , 2: n^x\,'   \rangle
= d_{\frac{\underline{n}^x-n^x}{2},\frac{\underline{n}^x\,'-n^x\,'}{2}}^{\frac{\underline{n}^x+n^x}{2}}(\theta)
\label{eq:generic_rot_OH}
\end{equation}
In the cylindrical basis of the 2D oscillator, one has  
\begin{multline}
|\underline{2}:(\underline{n}',\underline{m}'),2:(n',m')\rangle   \\
= R_{12}^{x}(\theta) R_{12}^{y}(\theta)|\underline{1}:(\underline{n},\underline{m}), 1:(n,m) \rangle .
\end{multline}
It is then useful to introduce the right and left annihilation operators
\begin{align}
&a_1^{\rm r}= \frac{1}{\sqrt{2}} \left(a_1^x-i a_1^y \right) \quad ; \quad 
a_1^{\rm l}= \frac{1}{\sqrt{2}} \left(a_1^x+i a_1^y \right) \\
&\underline{a}_1^{\rm r}= \frac{1}{\sqrt{2}} \left(\underline{a}_1^x-i \underline{a}_1^y \right) 
\quad ; \quad 
\underline{a}_1^{\rm l}= \frac{1}{\sqrt{2}} \left(\underline{a}_1^x+i \underline{a}_1^y \right)
\end{align}
and can use the identity
\begin{multline}
(\underline{a}_1^x)^\dagger a_1 -  a_1^\dagger \underline{a}_1^x + (\underline{a}_1^y)^\dagger {a}_1^y - ({a}_1^y)^\dagger \underline{a}_1^y\\
= (\underline{a}_1^{\rm d})^\dagger a_1^{\rm d} - (a_1^{\rm d})^\dagger \underline{a}_1^{\rm d} +    (\underline{a}_1^{\rm g})^\dagger a_1^{\rm g} - (a_1^{\rm g})^\dagger \underline{a}_1^{\rm g}
\end{multline}
to obtain
\begin{multline}
|\underline{2}:(\underline{n}',\underline{m}'),2:(n',m')\rangle   \\
= R_{12}^{\rm r}(\theta) R_{12}^{\rm l}(\theta)|\underline{1}:(\underline{n},\underline{m}),1:(n,m)\rangle
\end{multline}
where the rotation operators are given by
\begin{equation}
R_{12}^{\rm r}(\theta)= e^{\theta ((\underline{a}_1^{\rm r})^\dagger a_1^{\rm r} - (a_1^{\rm r})^\dagger \underline{a}_1^{\rm r})/2}
\end{equation}
and an analogous definition for ${R_{12}^{\rm l}(\theta)}$. The eigenvalues of the number operators ${(a_i^{\rm r})^\dagger a_i^{\rm r}}$ and ${(a_i^{\rm l})^\dagger a_i^{\rm l}}$ are related to the cylindrical quantum numbers by:
\begin{align}
&n_1^{\rm r}  = n + m  \quad ; \quad n_1^{\rm l} = n - m\\
&\underline{n}_1^{\rm r}  = \underline{n} + \underline{m}  \quad ; \quad \underline{n}_1^{\rm l} = \underline{n}-\underline{m}\\
&n_2^{\rm r}  = n' + m'  \quad ; \quad n_2^{\rm l} = n'-m'\\
&\underline{n}_2^{\rm r}  = \underline{n}' +\underline{m}' \quad ; \quad \underline{n}_2^{\rm l} = \underline{n}'-\underline{m}'
\end{align}
Finally, using the generic result of Eq.~\eqref{eq:generic_rot_OH} for a 1D oscillators, the desired scalar products can be thus expressed in terms of the Wigner d-matrix with
\begin{multline}
\langle \underline{1}:(\underline{n},\underline{m}),1:(n,m) |\underline{2}:(\underline{n}',\underline{m}'),2:(n',m')\rangle \\
=\langle \underline{1}:\underline{n}_1^{\rm r},1:n_1^{\rm r} |\underline{2}:\underline{n}_2^{\rm r},2:n_2^{\rm r}\rangle\\
\times \langle \underline{1}:\underline{n}_1^{\rm l},1:n_1^{\rm l} |\underline{2}:\underline{n}_2^{\rm l},2:n_2^{\rm l}\rangle.
\end{multline}

\end{document}